\documentclass[]{article}

\usepackage{color}
\usepackage[usenames,dvipsnames,svgnames,table]{xcolor}
\usepackage{graphicx}

\small

\def\mstar  {$M_{\star}$}
\def\macc   {$\dot{M}_{\rm acc}$}
\def\lacc   {$L_{\rm acc}$}

\def\msun {$M_{\odot}$}

\def\rstar {$R_\star$}

\def\mdisk {$M_{\rm disk}$}

\begin{document}

\twocolumn

\begin{center}
\fboxrule0.02cm
\fboxsep0.4cm
\fcolorbox{Brown}{Ivory}{\rule[-0.9cm]{0.0cm}{1.8cm}{\parbox{7.8cm}
{ \begin{center}
{\Large\em Perspective}

\vspace{0.5cm}

{\Large\bf Accretion onto young stars: the key to disk evolution}

\vspace{0.2cm}

{\large\em Carlo F. Manara}

% pls do not add affiliation

\vspace{0.5cm}

\centering
\includegraphics[width=0.19\textwidth]{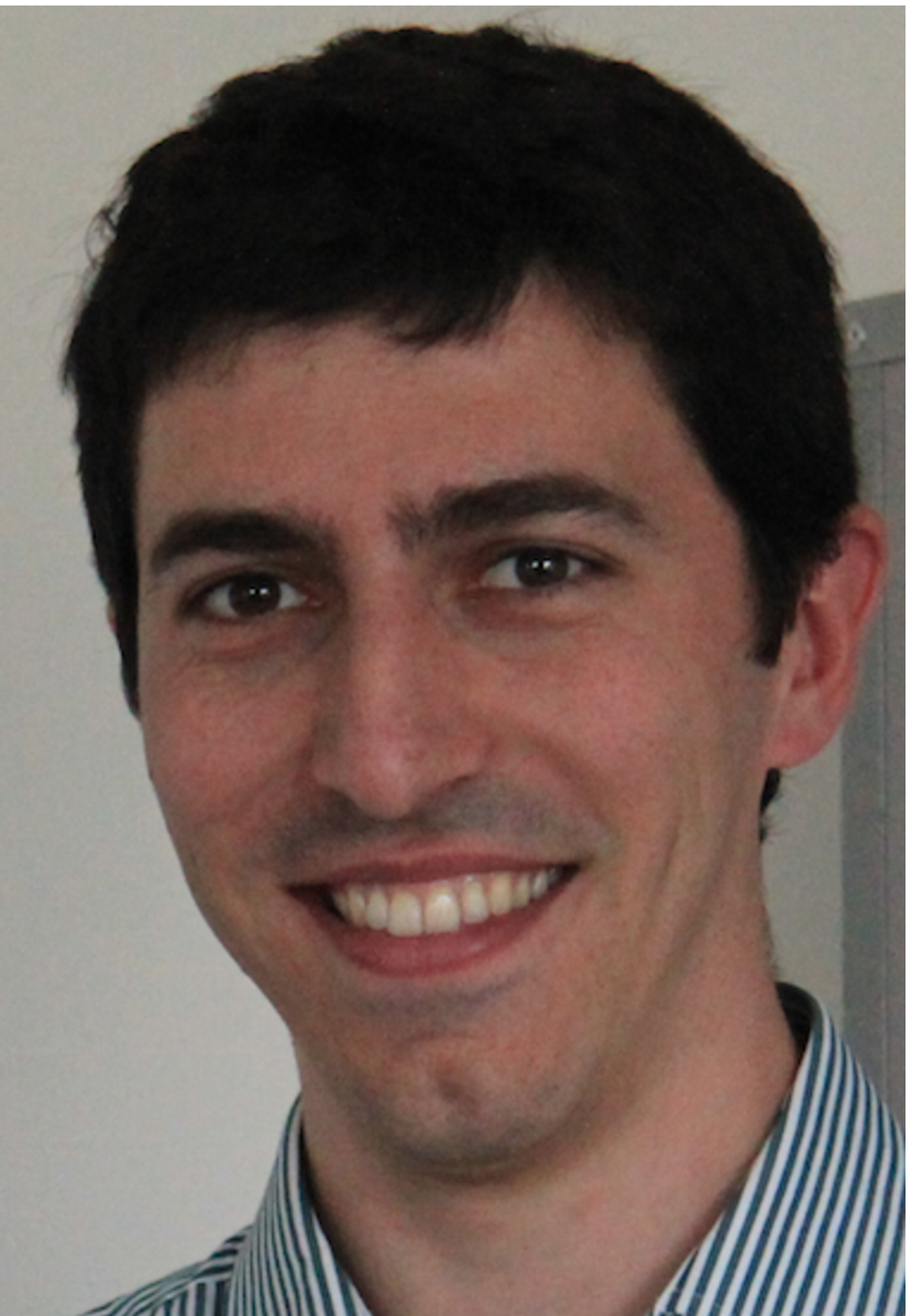}
\end{center}
}}}
\end{center}

\normalsize

%FROM BO: The idea with the articles is to give an overview of a certain subject from the point of view of a person who has been deeply involved. Hence it focuses to a large extent on that person's work, but put into a larger context, with an extensive introduction explaining briefly the history of the subject with key references. Opposing views are welcome as long as they are contrasted so their differences are clearly explained. Evidently your X-shooter data provide an excellent basis to discuss a number of issues in T Tauri stars, and you have free hands to choose the theme of the article. I would be glad if you would consider to write such an article for the Perspective series, summarizing your results from the X-shooter observations, and tying them into the larger picture of accretion in star formation.  

\section{Introduction}

The early stages of the pre-Main-Sequence (PMS) evolution of young
stars are characterized by a substantial interaction between the
central star and the surrounding disk. Among the processes that take place,
the accretion of material from the disk onto the star is a crucial
one, as it traces the disk evolution and, in turn, affects how planets are
formed and evolve in disks. Even if accretion has been studied for
many years, it is still the focus of a large body of work, as we
do not fully understand how it happens (see e.g., Hartmann et al. 2016
for an extended review). %This process is still crucial to study, both
%in order to better understand how it happens, and to trace the disk
%evolution processes to understand to how planets are formed in
%protoplanetary disks.

The current paradigm describing how the material is accreted from the disk onto the star is the magnetospheric accretion scenario. In this scenario the material is accreted onto the central star along the stellar magnetic field lines, assumed to be mostly dipolar. This paradigm was proposed almost 30 years ago to explain the ultraviolet excess and strong emission lines observed in the spectra of young stars (e.g., Camenzind 1990, K\"onigl 1991, Calvet \& Hartmann 1992). From then on, the comparison of more advanced observations with modeling of the process has confirmed magnetospheric accretion to be the way material is accreted onto young stars (e.g., Calvet \& Gullbring 1998, Muzerolle et al. 2003), at least in objects with stellar mass ($M_\star$) smaller than $\sim 2 M_\odot$, but also in some Herbig Ae/Be stars (e.g., Calvet et al. 2004, Mendigutia et al. 2011). Studies of the variability pattern of young stars are also in line with this paradigm (e.g., Venuti et al. 2015, Costigan et al. 2014). 
% Large efforts in measuring the strength and morphology of magnetic fields of young stars (e.g. Donati et al. 2007, 2010a, 2013; Hussain et al. 2009) are measuring strong multipolar components of the magnetic field, but it is not excluded that the actual strength of the bipolar component may be higher (e.g., Chen \& Johns-Krull 2013, Hussain \& Alecian 2014). Future studies, for example with CFHT/SPIRou, will definitely set what are the typical properties of magnetic fields of young stars.
A significant effort is going into measuring the strengths and large-scale topologies of magnetic fields at the surfaces of young stars (e.g., Donati et al. 2007, 2010a, 2013; Hussain et al. 2009, future: CFHT/SPIRou). These studies report strong (kG) multipolar components of the magnetic field but cannot exclude even stronger fields, particularly on small scales (e.g., Chen \& Johns-Krull 2013, Hussain \& Alecian 2014).
% Future studies, for example with CFHT/SPIRou and VLT/CRIRES+, will
% help to provide a more complete picture of the range of field
% strengths in young stars and their corresponding large-scale
% topologies.

Overall, it is now considered fair to assume that, in young stars, the excess continuum and line emission with respect to the photospheric and chromospheric one is (mainly) due to accretion. This excess emission can be converted into accretion luminosity (\lacc), and then mass accretion rates (\macc) by knowing \mstar \ and the stellar radius (\rstar). Blue spectra of young stars, ideally near-ultraviolet (NUV) spectra taken with the Hubble Space Telescope (HST, e.g., Herczeg \& Hillenbrand 2008, Ingleby et al. 2013) are ideal to trace excess continuum emission. Optical and near-infrared spectra are also used to measure the fluxes of hydrogen, helium, and calcium emission lines produced by accretion (e.g., Calvet \& Hartmann 1992, Natta et al. 2004, 2006, Mohanty et al. 2005, Antoniucci et al. 2011, Fang et al. 2013).
%This excess emission is detectable either using blue spectra of young stars, ideally near-ultraviolet (NUV) spectra taken with the Hubble Space Telescope (HST, e.g., Calvet+, Valenti et al. 1993, Ingleby et al. 2013, HH08) or optical and near-infrared spectra covering hydrogen, helium, and calcium emission lines (e.g., Calvet+, Natta et al. 2004, 2006, Mohanty et al. 2005, Fang et al. 2013, Antoniucci et al. 2011, Biazzo et al.+, Costigan+ etc). 
Photometry in the $U$-band (e.g., Gullbring et al. 1998, Robberto et al. 2004, Sicilia-Aguilar et al. 2010, Rigliaco et al. 2011, Manara et al. 2012, Venuti et al. 2014) or in narrow band filters, such as filters centered on the H$\alpha$ line (e.g., De Marchi et al. 2010, Kalari et al. 2015), is also a powerful way to measure \lacc. While photometry provides access to large samples with a limited telescope time effort, spectroscopy allows us to firmly determine the accretion and stellar properties.
\begin{figure*}[ht!]
\centering
\includegraphics[width=0.9\textwidth]{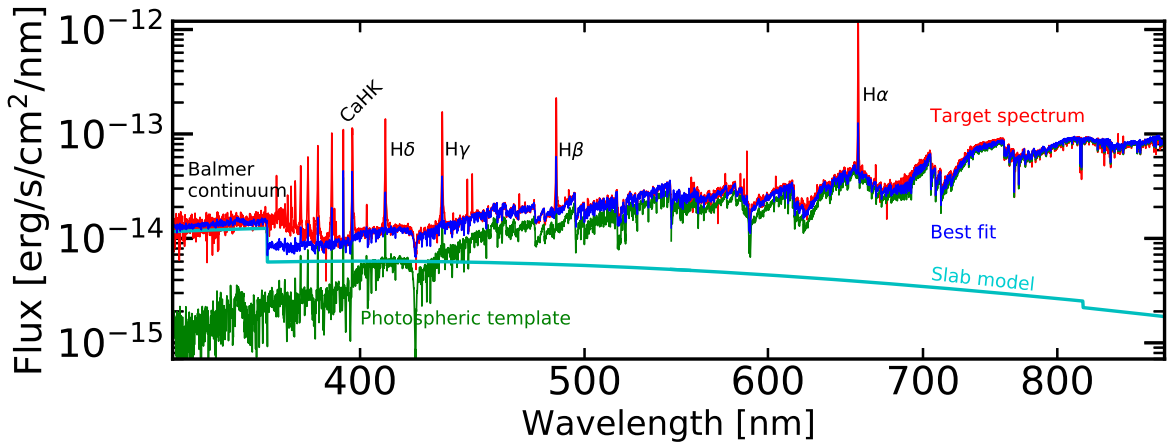}
% FIGURES MUST BE PS OR EPS
\vspace{-0.7cm}
\caption{Fit of the target Ass Cha T 2-10 (adapted from Manara et al. 2017a). The target spectrum, corrected for extinction ($A_V$=1.1 mag), is shown in red. The best fit of the continuum emission is shown in blue, and it is the sum of the photospheric template (green) and of the accretion shock model (slab model, cyan). }
\label{fig:fitxs}
\vspace{-0.5cm}
\end{figure*}

One big elephant in the room in this context is which process allows the material to lose angular momentum and therefore accrete \textit{through} the disk. This issue is key to determine how disks evolve and how planets form. Indeed, the processes proposed to explain the evolution of disks and the transport of material through the disk predict very different surface density distributions at a given time. These differences strongly impact how and where planets are formed in the disk, and how they migrate (e.g., Thommes et al. 2008, Mordasini et al. 2012, Bitsch et al. 2015, Morbidelli \& Raymond 2016). The historically favored process to describe the redistribution of material and angular momentum in disks is viscous accretion (e.g., Lynden-Bell \& Pringle 1974, Hartmann et al. 1998), usually coupled to processes such as internal photoevaporation driven by high-energy UV- and/or X-ray radiation, crucial to explain disk dispersal (e.g., Alexander et al. 2014, Gorti et al. 2016, Ercolano \& Pascucci 2017). 
Recently, more interest is being devoted to study magnetic disk winds (e.g., Armitage et al. 2013, Bai 2016, Ercolano \& Pascucci 2017) and low viscosity hydrodynamic turbulence (e.g., Hartmann et al. 2017), which are other ways to explain the removal of angular momentum.
%Recently, more interest is being devoted to trying to explain removal of angular momentum through magnetic disk winds (e.g., Armitage et al. 2013, Bai 2016, Ercolano \& Pascucci 2017), or through low viscosity hydrodynamic turbulences (e.g., Hartmann et al. 2017). 
These processes predict how the mass accretion rate scales with age, with stellar mass (see Sect.~3), and with the disk mass (see Sect.~4). Therefore, determining stellar, accretion, and disk properties in large samples of young stellar objects (see Sect.~2) is key to determine which process regulates disk evolution, hence constraining planet formation.

%In the following I will discuss how accretion rates are now measured by using broad-band medium-resolution flux-calibrated spectra (Section~2), and how these values are observed to scale with the stellar mass (Section~3) and with the disk mass (Section~4), while comparing these findings with our current theoretical explanations. Finally, I wrap up the conclusions and future wishes in Section~5. 

\section{Accretion rate estimates in the VLT/X-Shooter era}

The spectral type of a star determines the overall shape of its
spectrum and the appearance and strength of absorption lines and
molecular bands. However, the presence of usually non-negligible
interstellar extinction, different from object to object, makes the
overall shape of the spectrum redder. In contrast, the emission
due to accretion gives rise to a strong blue continuum excess. One
needs to take into account all these aspects simultaneously to be able
to overcome the degeneracies and derive the stellar and accretion
parameters of a young star. This is possible only by combining good
quality broad-band flux-calibrated spectra (e.g., Manara et al. 2013b,
Herczeg \& Hillenbrand 2014), with a modeling of the various
components. In Manara et al. (2013b) we showed how to overcome
degeneracies between the parameters with the method I will describe in
this section. We analyzed two objects in the Orion Nebula Cluster,
previously considered to be older than 30 Myr based on their location
on the HR Diagram. Our analysis method showed that they have stellar
parameters typical of $\sim$2-3 Myr old objects, much more in line
with the rest of the population of the cluster. The previous estimates
were affected by a degeneration between $A_V$ and \lacc.

A key ingredient in the development of our new analysis method has been the advent of the spectrograph X-Shooter (Vernet et al. 2011), a second generation instrument on the ESO Very Large Telescope (VLT). %, has revolutionized the way we analyze spectra of young stars. 
This instrument allows one to obtain flux-calibrated spectra covering simultaneously the wavelength interval from $\sim$300 nm to $\sim$2.5 $\mu$m with a resolution $R\sim$ 5000-18000. This implies covering simultaneously the spectral region of the Balmer continuum, particularly sensitive to excesses due to accretion, several emission lines tracing both accretion and winds, and multiple absorption features needed to determine the stellar properties of the target and the veiling due to accretion. Given these great potentials, a significant fraction of the Italian X-Shooter guaranteed time (GTO) was devoted to collect spectra for a large sample of young stars both with and without disks to determine their stellar and accretion properties (Alcal\'a et al. 2011). 

During my PhD I used these GTO spectra and I developed with my colleagues a self-consistent method to obtain the spectral type and stellar luminosity, reddening, and the accretion luminosity of young stars (Manara et al. 2013b). The method is based on finding the best fit model in a grid by minimizing a $\chi^2_{\rm like}$ distribution computed in various continuum regions of the spectra from the Balmer continuum ($\lambda\sim330$ nm) to the red part of the optical spectrum ($\lambda\sim$ 720 nm). The grid of models spans a set of photospheric templates of young stars observed by us with X-Shooter, a number of extinction ($A_V$) values, and models of the continuum excess due to accretion (a hydrogen slab model, as done by, e.g., Valenti et al. 1993 and Herczeg \& Hillenbrand 2008). The photospheric templates we use are observed X-Shooter spectra of non accreting young stars with spectral types between G5 and M9.5 (Manara et al. 2013a, 2017b), the best possible templates to reproduce the expected photospheric emission of young stars. %and were recently augmented by some earlier type objects with spectral type from G5 to K6, and with a few more late type templates with spectral type from M6.5 to M8 (Manara et al. 2017b). 
Our modeling allows us to reproduce the continuum emission from the target spectra of young stars from $\lambda\sim$ 300 nm up to $\lambda\sim$ 900 nm (Fig.~\ref{fig:fitxs}), and to derive stellar parameters and mass accretion rates with typical uncertainties of $<$0.45 dex in \macc \ (e.g., Alcal\'a et al. 2014). 
%Furthermore, in Manara et al. (2013b) we showed how to overcome degeneracies between the parameters. In that work we analyzed two objects in the Orion Nebula Cluster, previously considered to be older than 30 Myr based on their location on the HR Diagram. Our analysis showed that they have stellar parameters typical of $\sim$2-3 Myr old objects, much more in line with the rest of the population of the cluster. The previous estimates were affected by a degeneration between $A_V$ and \lacc. %, that was overcomed by our method.

More recently we have collected larger samples of VLT/X-Shooter spectra of young stars in different star forming regions in addition to the GTO data. We aimed at covering objects with a large range of \mstar \ and located in different regions. 
By analyzing the spectra with the method described above we have been able to study how the mass accretion rates scale with the stellar mass, and, combining these surveys with the results from ALMA surveys, how \macc \ scales with the disk mass, as I describe in the following sections. 
I only mention that these spectra are useful for many other goals, including calibrating the relation between emission lines and accretion luminosity (Alcal\'a et al. 2014, 2017), studies of winds and jets in young stars (Natta et al. 2014, Nisini et al. 2017), of photospheric properties (Stelzer et al. 2013, Frasca et al. 2017) and of metallicity of young stars (Biazzo et al. 2017). 
We used our spectra of non-accreting young stars to determine how chromospheric emission makes the detection of low \macc \ challenging (Manara et al. 2013a, 2017b), giving a quantitative value in line with previous studies (e.g., Sicilia-Aguilar et al. 2010, Ingleby et al. 2011). We have studied individual objects, for example DR Tau, to determine the impact of accretion variability on the chemical evolution of its disk (Banzatti et al. 2014). We have also studied the accretion properties of transition disks, finding that they have in many cases accretion rates as high as those of full disks (e.g., Manara et al. 2014), in line with e.g., Espaillat et al. (2014) or Najita et al. (2015). 
Finally, I would like to stress that the method described here is also applicable to other broad-band flux-calibrated spectra, such as LBT/MODS spectra, as we showed in the analysis of the EXor V1118 Ori (Giannini et al. 2017).

\section{The relation between stellar and accretion properties}
Our first large sample of stars with disks in a single star forming region, Lupus, was collected during the GTO observations. The analysis of this incomplete sample showed a simple power-law relation between \macc \ and \mstar \ with a spread $<$1 dex, much narrower than in previous studies (Alcal\'a et al. 2014). The other initial and still incomplete samples of X-Shooter spectra of young stars with disks in the $\sigma$-Orionis and $\rho$-Ophiuchus regions were analyzed by Rigliaco et al. (2012) and Manara et al. (2015), respectively. These datasets supported the idea that the power-law relation between \macc \ and \mstar \ found in the first sample of targets in the Lupus region is valid from brown dwarfs up to $\sim$1 \msun \ with a scatter smaller than $\sim$1 dex. 

\begin{figure}[b!]
\centering
\includegraphics[width=0.5\textwidth]{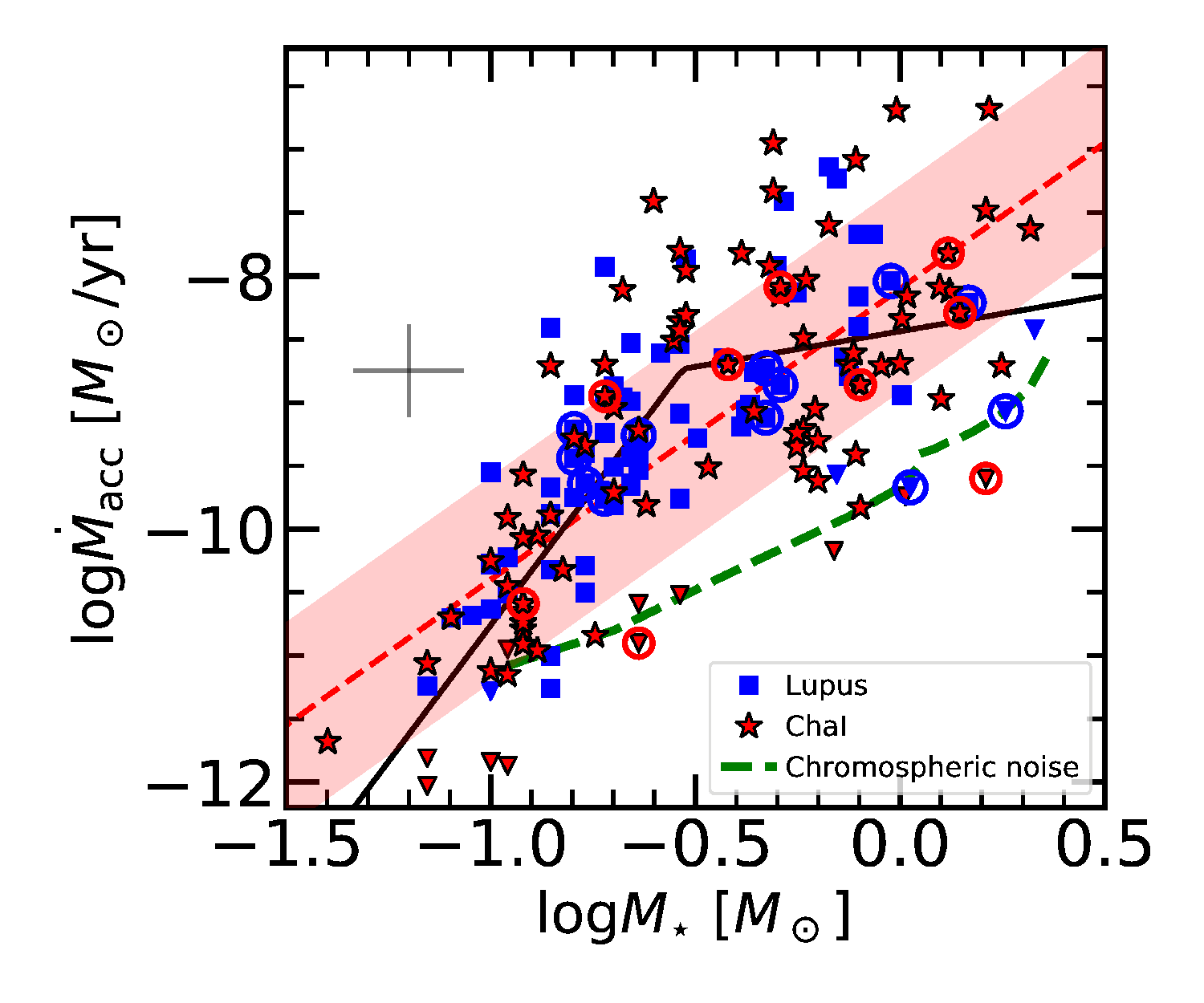}
\vspace{-5mm}
% FIGURES MUST BE PS OR EPS
\caption{Mass accretion rates vs stellar mass for the samples of stars with disks in the Lupus and Chamaleon~I regions (Alcal\'a et al. 2014, 2017; Manara et al. 2016a, 2017a). Transition disk objects are circled, downwards triangles are targets with accretion rates compatible with chromospheric emission, which is expected to be at about the location of the dashed green line at this age (Manara et al. 2013a, 2017b). The fits for the Chamaeleon~I sample are shown with a red dashed line (single power-law) and a black solid line (double power-law). }
\label{fig:macc_mstar}
\end{figure}
However, when we recently collected and analyzed the almost complete samples of X-Shooter spectra of young stars with disks in the Chamaeleon~I and Lupus star forming regions (Fig.~\ref{fig:macc_mstar}, Manara et al. 2016a, 2017a; Alcala et al. 2017), a more elaborate picture emerged: 
%First of all, the scatter of \macc \ at \mstar$>$0.3\msun \ is as large as $\sim$2 dex, and possibly smaller for smaller \mstar. 
the \macc-\mstar \ relation in each of the two regions is better represented with a slightly higher statistical confidence by a double power-law fit, with a steeper relation at low \mstar, than by a simple power-law. This double power-law behaviour, tentatively already observed also by Fang et al. (2013) and Venuti et al. (2014) in other complete samples in other regions, is quantitatively similar to what Vorobyov \& Basu (2009) suggested as a result of two different accretion regimes at different \mstar. However, it could also result from a faster evolution of the accretion rates for lower mass stars, as proposed by Rigliaco et al. (2011), Manara et al. (2012), or Fang et al. (2013). On the other hand, the data can also be fitted with a single power-law with exponent $\sim$2. This single power-law is possibly explainable by different theories, such as the result of particular initial conditions (Alexander \& Armitage 2006; Dullemond et al. 2006), of Bondi-Hoyle driven accretion (Padoan et al. 2005), or, if the slope is $\sim$1.7, of an evolution driven by X-ray photoevaporation (Ercolano et al. 2014). 
An intriguing feature of these sets of data is the lack of low accretors at \mstar$\sim$0.3 \msun \ (Manara et al. 2017a). This could be a consequence of the fast dispersal of disks due to photoevaporation (see also Sicilia-Aguilar et al. 2010). 
Finally, the objects with a transition disk are well mixed in this plot with objects hosting a full disk, although none of the strongly accreting objects hosts a transition disk, and some transition disk objects appear to be very low accretors, or non-accretors. 

One still open question, from an observational point of view, is how this picture evolves with time. The two regions we have been studying with X-Shooter, Lupus and Chamaeleon~I, have similar ages, and the age differences of targets within these regions are subject to large uncertainties (e.g., Soderblom et al. 2014). It is therefore mandatory to explore this relation in younger and older regions to further constrain the theoretical scenarios.

\section{Combining stellar, accretion, and disk properties}

\begin{figure}[h!]
\centering
\includegraphics[width=0.5\textwidth]{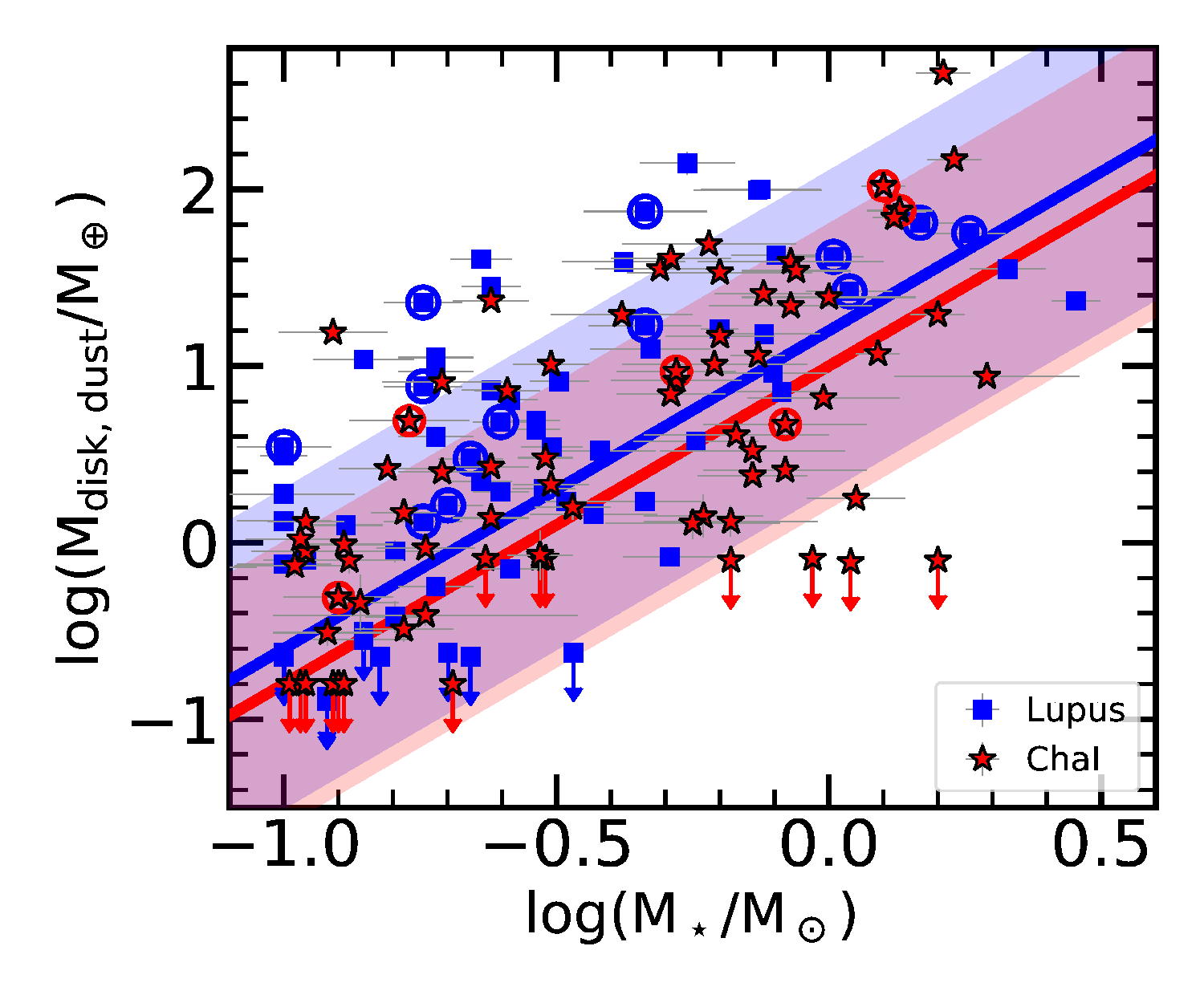}
% FIGURES MUST BE PS OR EPS
\caption{Disk mass vs stellar mass for the samples in the Lupus and Chamaeleon~I regions (Ansdell et al. 2016, Pascucci et al. 2016). Symbols as in Fig.~2. The stellar masses are derived from the stellar parameters derived using the X-Shooter spectra by Manara et al. (2016a, 2017a) and Alcal\'a et al. (2014, 2017). }
\label{fig:mstar_mdisk}
\end{figure}

The combination of precise values of \mstar \ and \macc \ with the measurements of disk masses from ongoing ALMA surveys in the Lupus (Ansdell et al. 2016) and Chamaeleon~I (Pascucci et al. 2016) star forming regions has allowed us for the first time to properly assess how the stellar and accretion properties of young stars scale with the disk masses. The relation between disk dust masses and stellar masses is found to be steeper than linear (Fig.~\ref{fig:mstar_mdisk}), in line with the higher number of massive planets found around more massive stars, and it becomes steeper with time (e.g., Ansdell et al. 2016, 2017, Pascucci et al. 2016, Barenfeld et al. 2016). It is worth noting that the transition disks happen to be among the most massive disks in the sample at all stellar masses, as noted by several authors (e.g., Van der Marel et al., subm.).

\begin{figure}[h!]
\centering
\includegraphics[width=0.5\textwidth]{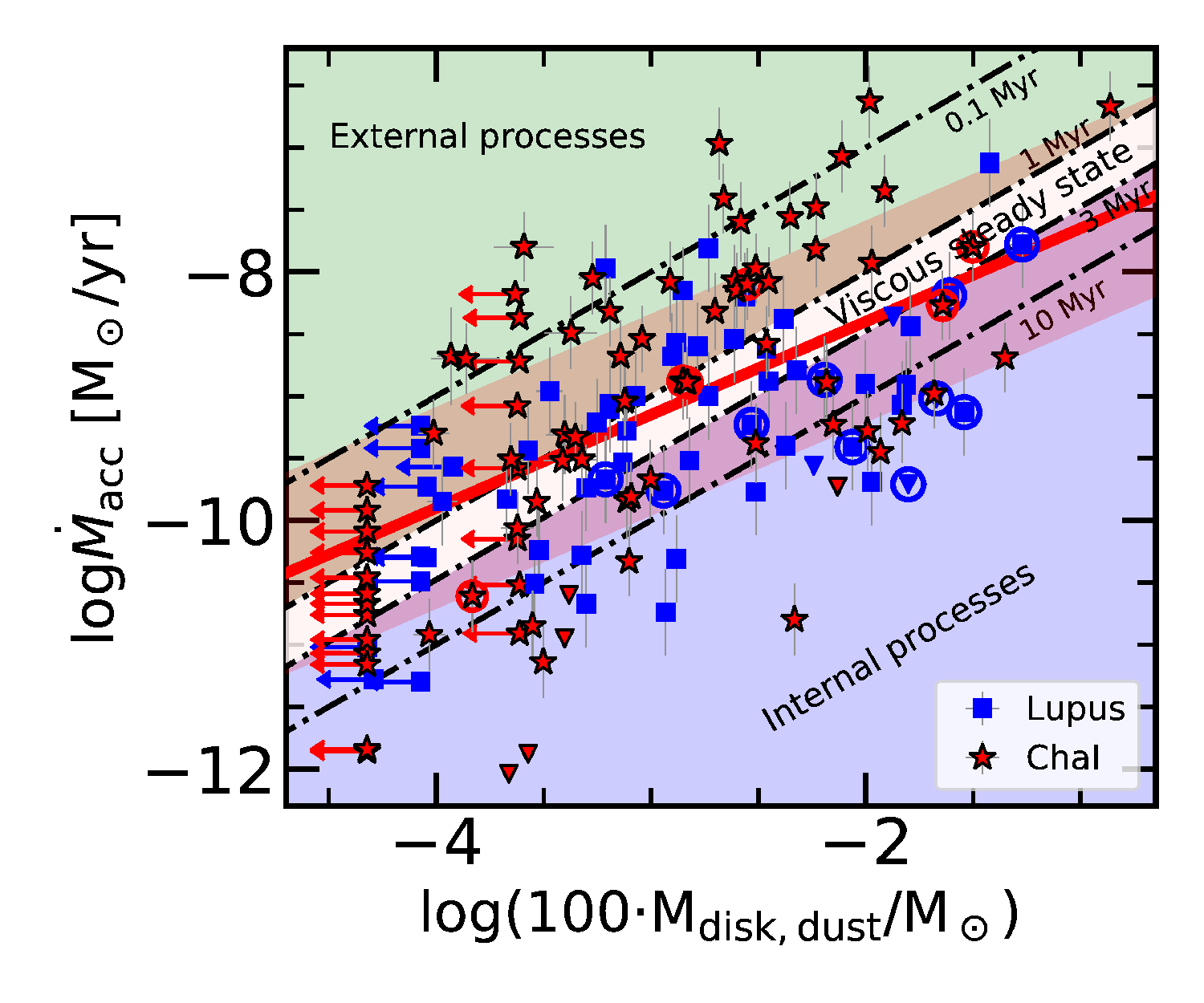}
% FIGURES MUST BE PS OR EPS
\vspace{-1.0cm}
\caption{Mass accretion rate vs disk mass for the samples in the Lupus and Chamaeleon~I regions (Manara et al. 2016b, Mulders et al. 2017). Symbols as in Fig.~2. The disk dust masses (Ansdell et al. 2016, Pascucci et al. 2016), are converted to total disk masses with a gas-to-dust-ratio of 100. Mass accretion rates are from Manara et al. (2016a, 2017a) and Alcal\'a et al. (2014, 2017). The red solid line is the best fit by Mulders et al. (2017). The black dot-dashed lines are values of \mdisk/\macc \ from 0.1 to 10 Myr. Colored regions are from Rosotti et al. (2017).}
\label{fig:macc_mdisk}
\end{figure}

The other important relation, the one between \macc\ and \mdisk, is
observed to be slightly shallower than linear in the Lupus and in the
Chamaeleon~I samples (Fig.~\ref{fig:macc_mdisk}, Manara et al. 2016b,
Mulders et al. 2017). Furthermore, the ratio \mdisk/\macc, expected for 
viscously evolving disks to
be approximately the age of the surrounding region 
(e.g., Jones et al. 2012, Rosotti et al. 2017), is found to be
$\sim$1-3 Myr for several disks in Lupus, as expected, when assuming a
gas-to-dust ratio of 100.  A very surprising finding is the fact that
the correlation is present when considering disk masses obtained from
(sub-)mm continuum measurements, thus tracing the dust content in
disks, and not when using disk gas masses from CO isotopologues (from
Ansdell et al. 2016 and Miotello et al. 2017).
% and using the dust masses to trace the disk masses, and usually much shorter when using the disk gas masses from CO. 
The lack of a correlation between \macc \ and the gas masses from CO observations, as well as the short inferred ratios \mdisk/\macc \ using CO gas masses, are further hints to the fact that CO gas masses are underestimated due to carbon depletion, as claimed by several independent studies (e.g., Bergin et al. 2014, Du et al. 2015, Kama et al. 2016, Miotello et al. 2017, Long et al. 2017).

Our finding of an almost linear correlation between \macc \ and the
disk (dust) mass is being used as a powerful test of our current
theories of disk evolution. Both Lodato et al. (2017) and Mulders et
al. (2017) have shown how the only way to reconcile the observed slope
and scatter around a linear relation with the predictions of viscous
evolution is to assume that the viscous timescale is as long as
$\sim$1 Myr, thus of the order of the age of the surrounding region.
%Although not theoretically impossible, such a long timescale is at odds with other findings, such as the observed timescale of dispersal of disks (e.g., Hernandez et al. 2007, Fedele et al. 2010, Bell et al. 2013).
Mulders et al. (2017) have explored a simple toy model of disks in which the mass accretion rate is independent of the disk mass, as expected if magnetic disk winds drive the evolution of disks. Such a model would lead to a distribution of \macc \ vs \mdisk \ very similar to the observed one. This is a very interesting result, but more realistic models of disk wind driven evolution are needed.

Rosotti et al. (2017) expanded the work of Jones et al. (2012) to explore the effect of several disk evolution processes on the \macc-\mdisk \ relation. This work shows that external photoevaporation (Clarke et al. 2007, Anderson et al. 2013) affects the disk mass so that the ratio \mdisk/\macc \ becomes much smaller than the age of the region. In contrast, internal processes, such as internal photoevaporation, planet formation, or dead zones, tend to produce ratios \mdisk/\macc \ larger than the age of the region. 
We combined ALMA measurements of disk masses and photometric estimates of \macc \ in the Orion Nebula Cluster (Mann et al. 2014, Eisner et al. 2016; Manara et al. 2012) and in the $\sigma$-Orionis cluster (Ansdell et al. 2017; Rigliaco et al. 2011), 
%In this context, the results obtained combining either disk masses from ALMA observations (Mann et al. 2014, Eisner et al. 2016) with $U$-band photometric \macc \ measurements (Manara et al. 2012) in the center of the Orion Nebula Cluster, or ALMA data (Ansdell et al. 2017) with $U$-band photometric measurements (Rigliaco et al. 2011) in the $\sigma$-Orionis cluster, 
and we showed that the observed ratios of \mdisk/\macc \ are very small, confirming that these disks are experiencing external photoevaporation (Rosotti et al. 2017, Ansdell et al. 2017).
% on this relation, showing that external processes, such as external photoevaporation (Clarke et al. 2007, Andersen et al. 2013), would modify the disk mass such that the ratio \mdisk/\macc \ would be much smaller than the age of the region. On the opposite, internal processes, such as internal photoevaporation, planet formation, dead zone, and more, would lead to values of the ratio \mdisk/\macc \ much larger than the disk age. 
Interestingly, the transition disk objects appear to be in the lower part of the distribution, meaning their ratios of \mdisk/\macc \ are in general larger than the age of the region, in line with previous results (e.g., Najita et al. 2015). In the theoretical framework discussed by Rosotti et al. (2017), this is in line with the fact that transition disk objects are experiencing internal processes, either planet formation or internal photoevaporation (e.g., Espaillat et al. 2014, Ercolano \& Pascucci 2017). 

Current observations thus imply that either disks evolve viscously but on a very long timescale and are dispersed by other mechanisms, or that several disks are already experiencing other processes than viscous evolution that are affecting their evolution. 

\section{What's next}

Our works based on VLT/X-Shooter spectra proved how crucial the use of broad-band flux-calibrated spectra is to overcome the degeneracies between stellar parameters, extinction, and the effect of accretion (e.g., Manara et al. 2013b), in particular when veiling due to accretion is substantial. We have also shown how the use of multiple emission line fluxes leads to estimates of the accretion luminosity consistent with that from the UV-excess (e.g., Alcal\'a et al. 2014). Again, this latter way to derive accretion rates relies on the availability of broad-band flux-calibrated spectra. Thus, it is crucial to continue to have access to instruments such as the VLT/X-Shooter, and in particular to cover the NUV Balmer continuum region to properly estimate the contribution of veiling in strongly accreting objects. Also HST (or its successor) is a key telescope, especially because it allows one to observe the FUV molecular and atomic gas lines emitted in the accretion shock and inner disk region (e.g., Ardila et al. 2013, France et al. 2014, Booth \& Clarke 2018). 

Furthermore, we have shown how important it is to survey complete
samples of young stars in the targeted regions, as incomplete samples
can lead to erroneous or only partially correct results in the same
way as inappropriate analysis methodologies.  The combination of
spectroscopic studies with surveys of disks is key to test disk
evolution mechanisms.
%, and the efforts from both communities should be coupled. One without the other would not be able to solve the puzzle of protoplanetary disk evolution, and thus planet formation. 
The relation between \macc \ and \mdisk, in particular, seems to be the most promising way to test models of angular momentum dispersal in disks.

Finally, the big question to address is the evolution of \macc \ with time. While Gaia will help us to lower our uncertainties on the ages of the regions (and maybe of individual objects?) by providing more stringent constraints on their distances and dynamical state (e.g., Voirin et al. 2017, Manara et al. 2017c), more effort should be put into covering all the stages of evolution of young stars. Obviously, it would be crucial to study older regions, where the disk evolution processes have shaped the disk properties. Moreover, the advent of JWST will allow us to finally be able to properly probe the earliest stages of disk evolution by measuring \macc \ in embedded objects, using the promising proxies of the mid-infrared emission lines (Salyk et al. 2013, Rigliaco et al. 2015). On this point, the still open question is how to determine the properties of the central forming star. Once this is sorted out, we will be able to finally determine how \macc \ evolves with time and constrain theories of disk evolution.

\footnotesize

{\it Acknowledgements:} I thank A. Natta, G. Hussain, J. Alcal\'a, and G. Rosotti for comments on a first draft of this article. I acknowledge support through the ESO Fellowship.
\newpage
{\bf References:}\\
Alcal\'a, J. M., Stelzer, B., Covino, E., et al. 2011, AN, 332, 242\\
Alcal\'a, J. M., Natta, A., Manara, C. F., et al. 2014, A\&A, 561, A2\\
Alcal\'a, J. M., Manara, C. F., Natta, A., et al. 2017, A\&A, 600, A20\\
Alexander, R. D., \& Armitage, P. J. 2006, ApJ, 639, L83\\
Alexander, R., Pascucci, I., Andrews, S., et al. 2014, PPVI, 475 \\
Antoniucci, S., García L\'opez, R., et al. 2011, A\&A, 534, A32\\
Anderson, K. R., Adams, F. C., \& Calvet, N. 2013, ApJ, 774, 9 \\
Ansdell, M., Williams, J. P., et al. 2016, ApJ, 828, 46\\
Ansdell M., Williams J. P., Manara C. F., et al. 2017, AJ, 153, 240\\
Ardila, D.R., Herczeg, G.J., Gregory, S., et al. 2013. ApJS, 207, 1\\ 
Armitage, P. J., Simon, J. B., \& Martin, R. G. 2013, ApJ, 778, L14 \\
Bai, X.-N. 2016, ApJ, 821, 80\\
Banzatti, A., Meyer, M. R., Manara, C.F. et al. 2014, ApJ, 780, 26\\
Barenfeld, S. A., Carpenter, J. M., et al. 2016, ApJ, 827, 142\\
Bergin, E. A., Cleeves, L. I., et al., 2014, FaDi, 168, 61\\ 
Biazzo, K., Frasca, A., Alcal\'a, J. et al. 2017, A\&A, 605, A66\\
Bitsch, B., Johansen, A., et al. 2015, A\&A, 575, A28\\ 
Booth, R.A., \& Clarke, C.J. 2018, MNRAS, 473, 757\\
Camenzind 1990, Reviews in Modern Astronomy, 3, 234-265\\
Calvet, N., \& Hartmann, L., 1992, ApJ, 386, 239\\
Calvet, N., \& Gullbring, E., 1998, ApJ, 509, 802\\
Calvet, N., Muzerolle, J., et al., 2004, AJ, 128, 1294\\
Chen, W., \& Johns-Krull, C.M., 2013, ApJ, 776, 113\\
Clarke, C. J. 2007, MNRAS, 376, 1350\\
Costigan, G., Vink, J.S., Scholz, A., et al. 2014, MNRAS, 440, 3444\\
De Marchi, G., Panagia, N., \& Romaniello, M., 2010, ApJ, 715, 1\\
Donati, J.-F., et al., 2007, MNRAS, 380, 1297\\
Donati, J.-F., et al., 2010a, MNRAS, 402, 1426\\
Donati, J.-F., et al., 2013, MNRAS, 436, 881\\
Du, F., Bergin, E. A., \& Hogerheijde, M. R. 2015, ApJL, 807, L32\\
Dullemond, C. P., Natta, A., \& Testi, L. 2006, ApJ, 645, L69\\
Eisner, J. A., Bally J. M., Ginsburg A., et al. 2016, ApJ, 826, 16\\
Ercolano, B., Mayr, D., Owen, J. E., et al. 2014, MNRAS, 178\\
Ercolano, B. \& Pascucci, I. 2017,  R. Soc. Open Sci., 4, 170114\\
Espaillat, C., Muzerolle, J., Najita, J., et al. 2014, PPVI, 497\\
Fang, M., Kim, J. S., van Boekel, R., et al. 2013, ApJS, 207, 5\\
France, K., Schindhelm, E., Bergin, E., et al. 2014, ApJ, 784, 127\\
Frasca, A., Biazzo, K., Alcal\'a, J. M., et al. 2017, A\&A, 602, A33\\
Giannini, T., Antoniucci, S., et al. 2017, ApJ, 839, 112\\
Gorti, U., Liseau, R., et al. 2016, Space Sci. Rev., 205, 125\\
Gullbring, E., Hartmann, L., et al. 1998, ApJ, 492, 323\\
Hartmann, L., Calvet, N., Gullbring, E., et al. 1998, ApJ, 495, 385\\
Hartmann, L., Herczeg, G., \& Calvet, N.\ 2016, ARA\&A, 54, 135 \\
Hartmann, L. \& Bae, J. 2017, MNRAS, in press, arXiv:1710.08718\\
Herczeg, G. J., \& Hillenbrand, L. A. 2008, ApJ, 681, 594\\
Herczeg, G. J., \& Hillenbrand, L. A. 2014, ApJ, 786, 97\\
Hussain, G.A.J., Collier Cameron, A. et al. 2009, MNRAS, 398, 189\\
Hussain, G.A.J. \& Alecian, E. 2014, IAUS, 302, 25\\
Ingleby, L., Calvet, N., Bergin, E., et al. 2011, ApJ, 743, 105\\
Ingleby, L., Calvet, N., Herczeg, G., et al. 2013, ApJ, 767, 112\\
Jones M. G., Pringle J. E., Alexander R., 2012, MNRAS, 419, 925\\
Kalari, V. M., Vink, J. S., et al. 2015, MNRAS, 453, 1026\\
Kama, M., Bruderer, S., et al. 2016, A\&A, 592, A83\\
K\"onigl 1991, ApJ, 370, L39\\
Lodato, G., Scardoni, C., Manara, et al. 2017, MNRAS, 472, 4700\\
Long, F., Herczeg, G.J., Pascucci, I. et al. 2017, ApJ, 844, 99\\
Lynden-Bell, D., \& Pringle, J. E. 1974, MNRAS, 168, 603\\
Manara, C.F., Robberto, M., Da Rio, N., et al. 2012, ApJ, 755, 154 \\
Manara, C.F., Testi, L., Rigliaco, E., et al. 2013a, A\&A, 551, A107 \\
Manara, C.F., Beccari, G., Da Rio, et al. 2013b, A\&A, 558, A114\\
Manara, C.F., Testi, L., Natta, A., et al. 2014, A\&A, 568, A18\\
Manara, C.F., Testi, L., Natta, A., et al. 2015, A\&A, 579, A66 \\
Manara, C.F., Fedele, D., Herczeg, G., et al. 2016a, A\&A, 585, A136\\
Manara, C.F., Rosotti, G., Testi, L., et al. 2016b, A\&A, 591, L3 \\
Manara, C.F., Testi, L., Herczeg, G., et al. 2017a, A\&A, 604, A127\\
Manara, C.F., Frasca, A., Alcal\'a, J.M. et al. 2017b, A\&A, 605, A86\\
Manara, C.F., Prusti, T. et al. 2017c, arXiv:1707.03179\\
Mann, R.K., Di Francesco, J., et al. 2014, ApJ, 784, 82\\
Mendigutia, I., Calvet, N., et al. 2011, A\&A, 535, 99\\
Miotello, A., van Dishoeck, E. F., et al. 2017, A\&A, 599, A113\\
Mohanty, S., Jayawardhana, R., \& Basri, G. 2005, ApJ, 626, 498\\
Morbidelli, A., \& Raymond, S. N. 2016, J. Geophys. Res., 121, 1962\\
Mordasini, C., Alibert, Y., Benz, W., et al. 2012, A\&A, 541, A97\\
Mulders, G., Pascucci, I., Manara, C.F. et al. 2017, ApJ, 847, 31\\
Muzerolle, J., Hillenbrand, L., Calvet, N. et al. 2003, ApJ, 592, 266\\
Najita, J.R., Andrews, S.M., Muzerolle, J., 2015, MNRAS, 450, 3559\\
Natta, A., Testi, L., Muzerolle, J., et al., 2004, A\&A, 424, 603\\
Natta, A., Testi, L., \& Randich, S. 2006, A\&A, 452, 245\\
Natta, A., Testi, L., Alcalá, J. M., et al. 2014, A\&A, 569, A5\\
Nisini, B., Antoniucci, S. et al. 2017, A\&A, in press, arXiv:1710.05587\\
Padoan, P., Kritsuk, A., Norman, M. L., et al. 2005, ApJ, 622, L61 \\
Pascucci, I., Testi, L., Herczeg, G. J., et al. 2016, ApJ, 831, 125\\
Rigliaco, E., Natta, A., Randich, S., et al. 2011, A\&A, 525, A47\\
Rigliaco, E., Natta, A., Testi, L., et al. 2012, A\&A, 548, A56\\
Rigliaco, E., Pascucci, I., Duchene, G., et al. 2015, ApJ, 801, 31\\
Robberto, M., Song, J., et al., 2004, ApJ, 606, 952\\
Rosotti G., Clarke C., Manara C.F. et al. 2017, MNRAS, 468, 1631\\
Salyk, C., Herczeg, G., Brown, J., et al. 2013, ApJ, 769, 21S\\
Sicilia-Aguilar, A., Henning, Th., Hartmann, L., 2010, ApJ, 710, 597\\
Soderblom, D. R., Hillenbrand, L. A., et al. 2014, PPVI, 219\\
Stelzer, B., Frasca, A., Alcal\'a, J. M., et al. 2013, A\&A, 558, A141\\
Thommes, E.W., et al. 2008, Science, 321, 814\\
Valenti, J. A., Basri, G., \& Johns, C. M. 1993, AJ, 106, 2024\\
Venuti, L., Bouvier, J., Flaccomio, E., et al. 2014, A\&A, 570, A82\\
Venuti, L., Bouvier, J., Irwin, J. et al. 2015, A\&A, 581, A66\\
Vernet, J., Dekker, H., D’Odorico, S., et al. 2011, A\&A, 536, A105\\
Voirin, J., Manara, C.F., \& Prusti, T., 2017, A\&A, in press, arXiv:1710.04528\\
Vorobyov, E. I., \& Basu, S. 2009, ApJ, 703, 922\\

\normalsize

\end{document}